\documentclass[12pt]{mn2e}
\usepackage{graphicx}
\usepackage{epsf}
\usepackage{lscape}
\usepackage{longtable}
\usepackage{rotating}
\usepackage{color}

\newcommand{\aap}{A\&A}

\newcommand{\apj}{ApJ}
\newcommand{\apjl}{ApJ}
\newcommand{\apjs}{ApJS}
\newcommand{\mnras}{MNRAS}
\newcommand{\aj}{AJ}
\newcommand{\apss}{Ap\&SS}
\newcommand{\pasp}{PASP}

\newcommand{\nat}{Nat}

\newcommand{\tr}{\textcolor{red}}

\begin{document}
\topmargin -0.5in 

\title[Single Colour Diagnostics of the Mass-to-light Ratio]{Single Colour Diagnostics of the Mass-to-light Ratio: Predictions from Galaxy Formation Models}

\author[Stephen M. Wilkins, et al.\ ]  
{
Stephen M. Wilkins$^{1}$\thanks{E-mail: stephen.wilkins@physics.ox.ac.uk}, Violeta Gonzalez-Perez$^{2}$, Carlton M. Baugh$^{2}$, \newauthor Cedric G. Lacey$^{2}$, Joe Zuntz$^{1,3}$ \\
$^1$\,University of Oxford, Department of Physics, Denys Wilkinson Building, Keble Road, Oxford, OX1 3RH, U.K. \\
$^{2}$\,Institute for Computational Cosmology, Department of Physics, University of Durham, South Road, Durham, DH1 3LE, U.K.\\
$^{3}$\,University College London, Department of Physics \& Astronomy, Gower Street, London, WC1E 6BT, U.K.\\
}
\maketitle 

\begin{abstract}
Accurate galaxy stellar masses are crucial to better understand the physical mechanisms driving the galaxy formation process. We use synthetic star formation and metal enrichment histories predicted by the {\sc galform} galaxy formation model to investigate the precision with which various colours $(m_{a}-m_{b})$ can alone be used as diagnostics of the stellar mass-to-light ratio. As an example, we find that, at $z=0$, the {\em intrinsic} (B$_{f435w}-$V$_{f606w}$) colour can be used to determine the intrinsic rest-frame $V$-band stellar mass-to-light ratio ($\log_{10}\Gamma_{V}=\log_{10}[(M/M_{\odot})/(L_{V}/L_{V\odot})]$) with a precision of $\sigma_{lg\Gamma}\simeq 0.06$ when the initial mass function and redshift are known beforehand. While the presence of dust, assuming a universal attenuation curve, can have a systematic effect on the inferred mass-to-light ratio using a single-colour relation, this is typically small as it is often possible to choose a colour for which the dust reddening vector is approximately aligned with the $(m_{a}-m_{b})-\log_{10}\Gamma_{V}$ relation. The precision with which the stellar mass-to-light ratio can be recovered using a single colour diagnostic rivals implementations of SED fitting using more information but in which simple parameterisations of the star formation and metal enrichment histories are assumed. To facilitate the wide use of these relations, we provide the optimal observer frame colour to estimate the stellar mass-to-light ratio, along with the associated parameters, as a function of redshift ($0<z<1.5$) for two sets of commonly used filters.
\end{abstract} 


\section{Introduction}

The accurate determination of the stellar masses of galaxies at different redshifts is crucial to improve our understanding of the process of galaxy formation and evolution.

The physical properties of galaxies, including the stellar mass-to-light ratio, are most often determined using spectral energy distribution (SED) fitting (see Walcher et al. 2011 for a recent overview), using broad-band photometry (e.g. Taylor et al. 2011), spectroscopy (e.g. Tojeiro et al. 2007, Pacifici et al. 2012), or spectral indicies (e.g. Kauffmann et al. 2003). The determination of the various properties (including the stellar mass-to-light ratio) involves the comparison of the observed SED with a library of templates. This comparison has been done either using a single best-fitting template (e.g. the template/model which minimises $\chi^{2}$) or carrying forward the full posterior probability distribution which contains all the information inferred about the stellar mass-to-light ratio (e.g. Kauffmann et al. 2003, Taylor et al. 2011.)

Each template/model is constructed using stellar population synthesis (SPS) modelling (e.g. Tinsley \& Gunn 1976, Bruzual \& Charlot 1993). Stellar evolution models are combined with a choice of initial mass function (IMF) to produce synthetic single-age stellar population (SSP) SED models as a function of age and metallicity (e.g. Bruzual \& Charlot 2003, Fioc \& Rocca-Volmerange 1997). By combining the SSP models with star formation and metal enrichment histories the {\em intrinsic} SED of a composite stellar population (CSP) can be assembled. By including absorption and emission due to dust and gas in the ISM and IGM (e.g. Madau et al. 1995) realistic SEDs of galaxies can be constructed. In its most basic incarnation, SED fitting assumes a template library built using a simple parameterisation of both the star formation and metal enrichment histories with some parameter(s) describing the effect of dust attenuation. The choice of star formation history (SFH) and metal enrichment history parameterisations varies between studies depending on the available photometry and/or spectroscopy. For example, at very-high redshift ($z>5$) where there is little high signal-to-noise photometry, a constant SFH is sometimes assumed (e.g. Gonzalez et al. 2011). However, a more common parameterisation is a decaying exponential: $\propto e^{-t/\tau}$ (which has two free-parameters, the age $t$ and characteristic timescale $\tau$) which is able to account for galaxies with little residual star formation. Further improvements include the addition of bursts (e.g. Kauffmann et al. 2003, Brinchmann et al. 2004, Salim et al. 2007, da Cunha et al. 2008, Walcher et al. 2009) or the truncation of star formation entirely. Further, some studies (e.g. Tojeiro et al. 2007) avoid the need for SFH and metal enrichment parameterisations altogether and instead construct CSPs from a series of discrete star formation episodes.

A different alternative to assuming a parameterisation is to build a training set (i.e. a set of observations with known properties). While it is possible to build an empirical training set in some cases (for example when the aim is to determine the redshift) in practice this is not possible for the stellar mass-to-light ratio as this cannot be directly measured (unlike the redshift). However, we can instead use galaxy formation models to produce physically motivated star formation and metal enrichment histories that can be used to generate what is essentially an {\em artificial} training set. This approach is of course only effective if the simulation is a good description of the real Universe. 

While it is possible to use an artificial training set as the basis of {\em full} SED fitting (i.e. using all available information) in this paper we focus on deriving a simple relationship between a single-colour and the stellar mass-to-light ratio. This is particularly useful for spectroscopic surveys of galaxies which may not have extensive multiband photometry. In addition, mass estimates based on a single-colour relation (unlike SED fitting with more than $2$-bands) are easily implemented by other studies permitting a straightforward comparison between samples. However, any empirical or theoretically defined single-colour - mass-to-light ratio relation is based on a specific population and may not be suitable for populations defined using an alternative criteria.

The use of a single-colour diagnostic has previously been considered in literature (at low redshift) and the strength of this approach has been demonstrated by, for example, Bell \& de Jong (2001) and more recently Taylor et al. (2011). For example, Taylor et al. (2011) compared stellar mass-to-light ratios derived from SED fitting using broad-band optical photometry collected by the Galaxy and Mass Assembly (GAMA, Driver et al. 2009, 2011) survey with the rest-frame $(g-i)$ colours finding a strong correlation.

It is also important to point out that stellar masses derived from any photometric method are sensitive to the choice of initial mass function (IMF). For example, stellar masses determined assuming a Salpeter (1955) IMF are, due to enhanced contribution of faint low-mass $<0.5\,{\rm M_{\odot}}$ stars, typically around $0.2\,{\rm dex}$ greater than those derived assuming a Chabrier (2003) IMF. This problem is exacerbated further if the IMF is not universal as suggested by several different studies (e.g.  Hoversten \& Glazebrook 2008, van Dokkum 2008, Wilkins et al. 2008a, Lee et al. 2009, Cappellari et al. 2012). Stellar mass estimates are also sensitive to uncertainties in our understanding of stellar evolution (e.g. Conroy et al. 2009) and thus vary depending on the choice of stellar population synthesis code (e.g. Maraston et al. 2003, Tonini et al. 2010). 

In this work we are interested firstly in determining the accuracy and precision with which a single colour can be used to infer the stellar mass-to-light ratio, the theory of which is described in Section \ref{sec:theory}. To achieve this we use the {\sc galform} galaxy formation model to generate realistic star formation and metal enrichment histories by following the physical processes that affect the baryonic component of the Universe. These are in turn used to produce synthetic {\em intrinsic} photometry for a large number of galaxies. The {\sc galform} model and filter sets utilised are described in \S\ref{sec:galform} and \S\ref{sec:filters} respectively. In Section \ref{sec:main} we demonstrate the precision with which a carefully chosen {\em intrinsic} colour can be used to determine the stellar mass-to-light ratio. We then investigate the effect of dust in \S\ref{sec:dust} and in Section \ref{sec:optimise} we determine, and present (along with the associated parameters), the optimal choice of colour with which to probe the mass-to-light ratio over $z=0.0\to 1.5$. In Section \ref{sec:sed_fitting} we compare the expected precision of a single-colour based diagnostic of the mass-to-light ratio with recent estimates of the accuracy and uncertainties of SED fitting. Finally in \ref{sec:c} we present our conclusions. Throughout this work we employ the $AB$ magnitude system (Oke \& Gunn 1983) and assume a Salpeter (1955) IMF, i.e. $\xi(m)={\rm d}N/{\rm d}m\propto m^{-2.35}$.

\section{The rest-frame optical colour as a diagnostic of the mass-to-light ratio}\label{sec:theory}

\begin{figure*}
\centering
\includegraphics[width=33pc]{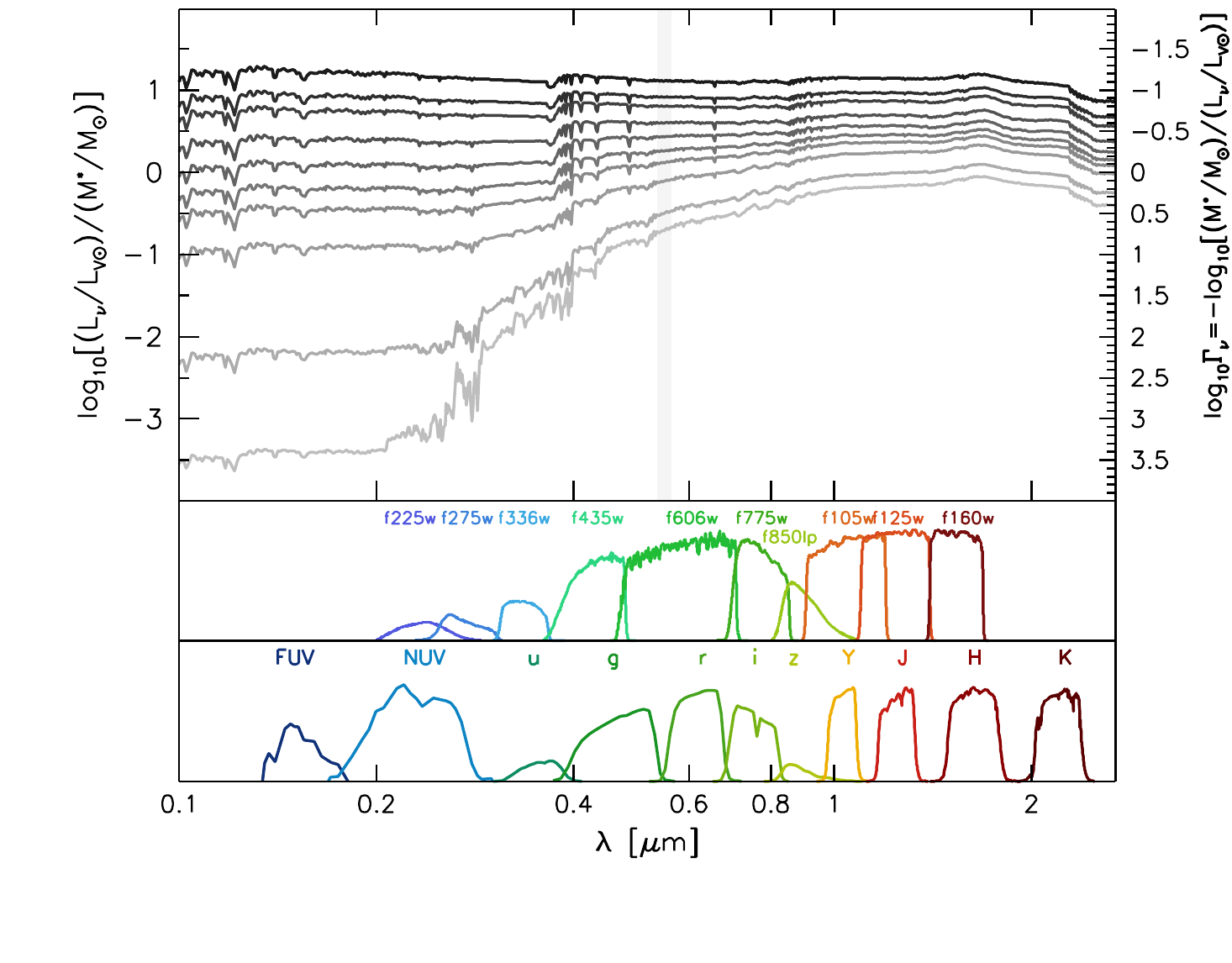}
\caption{{\em Upper panel}: The specific SED (i.e. per unit {\em current} stellar mass) assuming an exponentially decaying star formation history ($\propto e^{-t/\tau}$) with $\tau=1\,{\rm Gyr}$ and $\log_{10}(t/{\rm Gyr})=-1.0\to 1.0$ in increments of $\Delta\log_{10}(t/{\rm Gyr})=0.2$ (plotted using black to light grey lines). The {\em right}-hand axis shows the corresponding mass-to-light ratio (normalised by the $V$-band solar luminosity and solar mass). {\em Lower panel}: Transmission functions of the {\em HST} ACS and WFC3 filters considered in this study.}
\label{fig:sed}
\end{figure*}

The mass-to-light ratio of a stellar population is driven predominantly by the distribution of stellar masses and ages; the {\em specific} (i.e. per unit mass) luminosity of high-mass stars is much larger than for low-mass stars. The distribution of stellar masses is in turn driven by the star formation history (SFH) and initial mass function. Galaxies with protracted but declining star formation histories will contain, relative to galaxies with substantial recent star formation, more low-mass, lower-luminosity stars due to their longer main-sequence lifetimes.

\begin{figure}
\centering
\includegraphics[width=20pc]{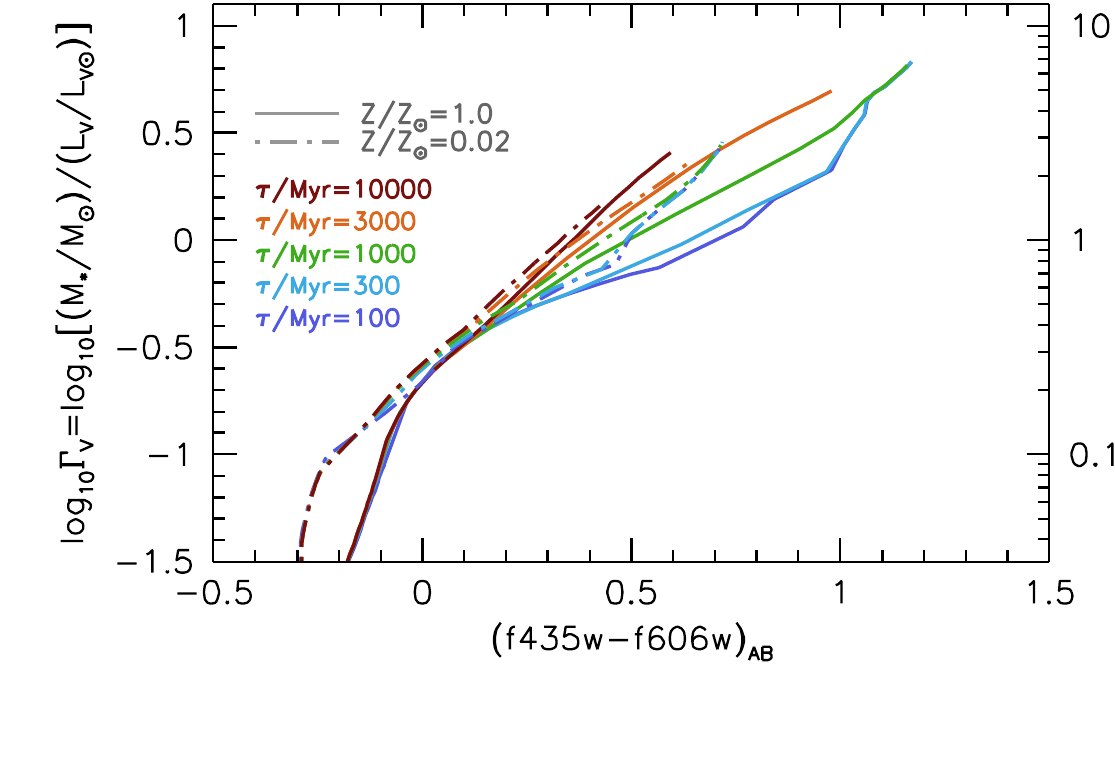}
\caption{Evolutionary tracks ($t=0\to t_{\rm uni}(z=0)$) of stellar populations in the {\em rest-frame} $(f435w-f606w)$ - $V$-band mass-to-light ratio plane, created assuming a Salpeter (1955) IMF, utilising the {\sc pegase.2} SPS model and assuming several exponentially declining star formation histories ($\psi\propto e^{-t/\tau}$ with $\tau/{\rm Gyr}\in \{0.1,0.3,1.0,3.0,10.0\}$) and two metallicities ($Z/Z_{\odot}=1$ and $Z/Z_{\odot}=0.02$) as indicated by the key.}
\label{fig:theory}
\end{figure}

Fig. \ref{fig:sed} shows the {\em specific} SED (per unit stellar mass) assuming an exponentially decaying star formation history ($\propto e^{-t/\tau}$) with $\tau=1\,{\rm Gyr}$ and $\log_{10}(t/{\rm Gyr})=-1.0\to 1.0$ in increments of $\Delta\log_{10}(t/{\rm Gyr})=0.2$. As the stellar population ages the specific SED decreases (the stellar mass-to-light ratio increases) and the rest-frame optical colour reddens. The correlation between stellar mass-to-light ratio and rest-frame optical colour can be seen more clearly in Fig. \ref{fig:theory} where evolutionary tracks ($t/{\rm Gyr}=0\to t_{\rm uni}(z=0)\equiv 13.5$) in the rest-frame {\em HST} ACS ($B_{f435w}-V_{f606w}$) colour - mass-to-light ratio plane are shown for various exponentially decaying star formation histories ($\psi\propto e^{-t/\tau}$ with $\tau/{\rm Gyr}\in \{0.1,0.3,1.0,3.0,10.0\}$) and two metallicities ($Z/Z_{\odot}=1$ and $Z/Z_{\odot}=0.02$) constructed using the {\sc pegase.2} SPS model (Fioc \& Rocca-Volmerange 1997, 1999) and assuming a Salpeter IMF.

For stellar populations with colours ($B_{f435w}-V_{f606w}$)$>0$\footnote{Colours $<0$ typically occur only for very young stellar populations.} the colour and mass-to-light ratio are correlated, i.e. for a given colour the range of stellar mass-to-light ratios is substantially smaller than the {\em full} range of mass-to-light ratios. While stellar populations with lower metallicities extend over a smaller range of mass-to-light ratios and colours, the mass-to-light ratio at the same colour is roughly similar, suggesting that using an optical colour to infer the mass-to-light is somewhat insensitive to the metal enrichment history.

However, it is difficult to derive an accurate single-colour mass-to-light ratio relation, together with an estimate of the uncertainty, from such simple and unrealistic star formation and metal enrichment histories. Instead it is more useful to use a galaxy formation model to produce a realistic library of galaxy star formation histories.

\section{Modelling Approach}

\subsection{Galaxy Formation Model}\label{sec:galform}

We predict the stellar masses and colours of galaxies at different redshifts in a $\Lambda$CDM universe using the {\sc galform} semi-analytical galaxy formation model initially developed by Cole et al. (2000). Semi-analytical models solve physically motivated differential equations to follow the fate of baryons in a universe in which structure grows hierarchically through gravitational instability (see Baugh 2006 for an overview of hierarchical galaxy formation models).

{\sc galform} follows the main processes which shape the formation and evolution of galaxies. These include: (i) the collapse and merging of dark matter haloes; (ii) the shock-heating and radiative cooling of gas inside dark matter haloes, leading to the formation of galaxy discs; (iii) quiescent star formation in galaxy discs; (iv) feedback from supernovae, from active galactic nuclei (AGN) and from photoionization of the intergalactic medium (IGM); (v) chemical enrichment of the stars and gas; (vi) galaxy mergers driven by dynamical friction within common dark matter haloes, leading to the formation of stellar spheroids, which also may trigger bursts of star formation; (vii) bursts of star formation in dynamically unstable discs. The end product of the calculations is a prediction for the number and properties of galaxies that reside within dark matter haloes of different masses. These properties include the star formation and metal enrichment histories.

In this paper we focus our attention on the Bower et al. (2006, B06) variant of {\sc galform}. Key features of the B06 model include (i) a time scale for quiescent star formation that scales with the dynamical time of the disk and which therefore changes significantly with redshift (see Lagos et al. 2011 for a study of different star formation laws in quiescent galaxies), (ii) bursts of star formation occur due to both galaxy mergers and when disks become dynamically unstable, and (iii) negative feedback from both supernovae and AGN (see Benson et al. 2003 for a discussion of the impact of feedback has on the luminosity function of galaxies). The onset of the AGN suppression of the cooling flow is governed by a comparison of the cooling time of the gas with the free-fall time for the gas to reach the centre of the halo. Cooling is suppressed in quasi-static hot haloes if the luminosity released by material accreted onto a central supermassive black hole balances the cooling luminosity (see Fanidakis et al. 2011 for a full description of black hole growth in the model). Bower et al. (2006) adopt the cosmological parameters of the Millennium Simulation (Springel et al. 2005), which are in broad agreement with constraints from measurements of the cosmic microwave background radiation and large scale galaxy clustering (e.g. Sanchez et al. 2009): $\Omega_{0}=0.25$, $\Lambda_{0} = 0.75$, $\Omega_{b}=0.045$, $\sigma_{8}=0.9$ and $h=0.73$, such that the Hubble constant today is $H_0=100\,h$ km$\,{\rm s}^{-1}$Mpc$^{-1}$. The B06 model parameters were fixed with reference to a subset of the available observations of galaxies, mostly at low redshift (see Bower et al. 2010 for a discussion of parameter fitting). This model successfully reproduces the galaxy stellar mass function up to $z=4.5$ and the properties of red galaxies at $0.7<z<2.5$ (Gonzalez-Perez et al. 2009, 2011). For further details we refer the reader to B06.

For the work presented here, we have implemented the B06 model using merger trees generated with the Monte Carlo algorithm introduced by Parkinson et al. (2008). We have also changed the default SSP used to produce the SEDs of galaxies, using here {\sc pegase.2} combined with a Salpeter (1955) IMF\footnote{Note that B06 used the SSPs from Bruzual \& Charlot (1999) combined with a Kennicutt (1983) IMF.}. A Salpeter IMF is chosen over the default (Kennicutt 1983) IMF due to its wide use in the literature. These changes have only a minor effect on the intrinsic distribution of colours and do not affect the optimum choice of colour with which to probe the stellar mass-to-light ratio. Assuming a Salpeter IMF however {\em increases} the stellar mass-to-light ratios (relative to a Kennicutt IMF) and thus affects the normalisation of any relation between colour and stellar mass-to-light ratio. The adoption of Salpeter IMF modifies the predicted luminosity function in the expected way, introducing a shift of about $0.4\,{\rm mag}$ in the $b$-band luminosity function at z=0 with respect to observations.

In this work, we uniquely use the unextincted magnitudes output by {\sc
galform}. The attenuation of starlight by dust is modelled in
{\sc galform} in a physically self-consistent way, based on the results of a radiative transfer calculation for a realistic geometry and predicted galaxy sizes and dust masses (Cole et al. 2000, Lacey et al. 2011, Gonzalez-Perez et al. 2012). Nevertheless, here we do not make use of this part of the modelling (see Mitchell et al., {\em in prep.}, for a discussion on the stellar mass estimate using the standard dust treatment in {\sc galform}). Instead, here we apply a stand-alone model for dust extinction using the Calzetti et al. (2000) reddening law (see \S\ref{sec:dust}), one of the most widely used techniques for analysing the dust content of observed galaxies. This allows us to demonstrate the effect that a universal attenuation curve has on the stellar mass estimation.

\subsubsection{Sample definition}\label{sec:sample_defn}

Throughout this work we consider a stellar mass limited sample, choosing $\log_{10}(M_{*}/M_{\odot})>9$\footnote{In \S\ref{sec:mass_cut} we also consider the effect on the recovered parameters of changing this mass cut finding only a small effect.}. This limit is chosen to roughly reflect the mass range accessible to typical HST surveys at $z=0.5-1.5$.

The merger trees used in the model are generated using a Monte Carlo approach, for which weights are given to the haloes quantifying their abundance based on the halo mass function. In turn, each galaxy has the associated weight of its host halo encoding the relative probability with which it would occur. Thus, a histogram of the stellar mass of each galaxy in the sample incorporating these weights will then reproduce the galaxy stellar mass function. We carry these weights through our entire analysis to ensure that the sample truly reflects a mass-limited criteria.

\subsubsection{The predicted distribution of mass-to-light ratios}\label{sec:galform.mtold}

\begin{figure}
\centering
\includegraphics[width=20pc]{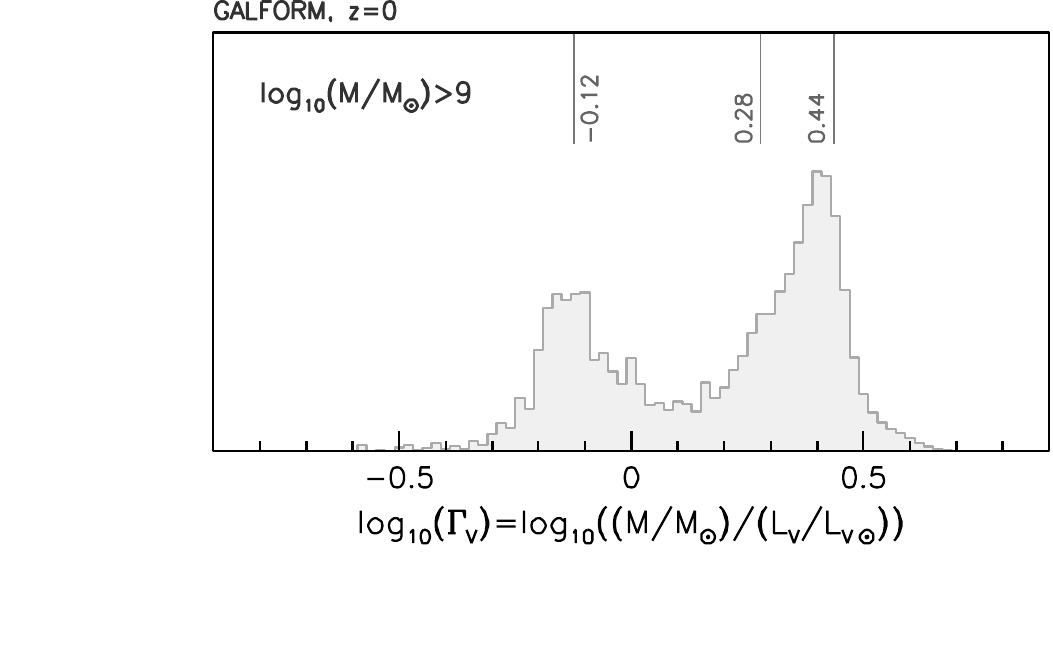}
\caption{The distribution of {\em intrinsic} rest-frame $V$-band mass-to-light ratios ($\Gamma_{V}$) for galaxies with $\log_{10}(M/M_{\odot})>9$ at $z=0$ predicted by the Bower et al. (2006) version of the {\sc galform} galaxy formation model assuming a Salpeter IMF and the {\sc pegase.2} SPS model. The vertical lines denote the $15.87^{\rm th}$, $50^{\rm th}$, and $84.13^{\rm th}$ percentiles of the distribution from left to right respectively.}
\label{fig:mtol}
\end{figure}

The distribution of intrinsic rest-frame $V$-band mass-to-light ratios ($\log_{10}\Gamma_{V}$) predicted by {\sc galform} is shown in Fig. \ref{fig:mtol}. The distribution spans a range of $\Delta\log_{10}\Gamma_{V}\simeq 0.61$ between the $10^{\rm th}$ to $90^{\rm th}$ percentiles and has a standard deviation of $\sigma_{\log_{10}\Gamma}=0.26$, though the distribution is non-gaussian and asymmetric. The bi-modality simply reflects the population of active and passive galaxies predicted by the model, corresponding to `red' and `blue' galaxies.

\subsection{Filter Sets}\label{sec:filters}

In this work we consider two sets of filters. The HST filter set employs a combination of {\em HST} ACS and WFC3 filters (WFC3.UVIS: UV$_{f225w}$, UV$_{f275w}$, U$_{f336w}$, ACS: B$_{f435w}$, V$_{f606w}$, $i_{f775w}$, $z_{f850lp}$, WFC3.NIR: Y$_{f105w}$, J$_{f125w}$, H$_{f160w}$). The choice of these filters is motivated by the existence of publically availability deep imaging in all or a significant combination of these filters. The GAMA filter set is based on the observations accessible to the GAMA survey and includes the Galaxy Evolution Explorer ({\em GALEX}, Martin et al. 2005) FUV and NUV, Sloan Digital Sky Survey (SDSS, Fukugita et al. 1996) $ugriz$ and the United Kingdom Infrared Telescope ({\em UKIRT}) Wide Field Camera (WFCAM, Casali et al. 2007) YJHK filters (Hewett et al. 2006). The filter transmission functions are shown in Fig. \ref{fig:sed}.

\section{Single-Colour Mass-to-light Ratio Relation}\label{sec:main}

\begin{figure}
\centering
\includegraphics[width=20pc]{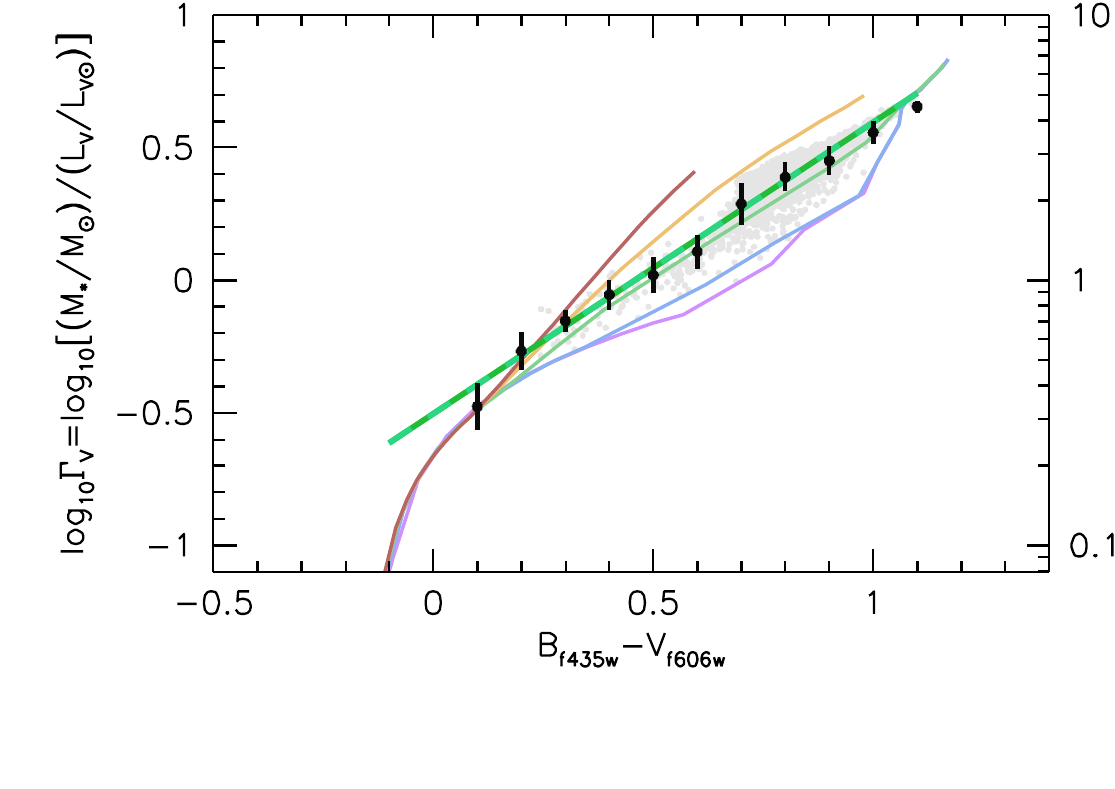}
\caption{The relationship between the (B$_{f435w}-$V$_{f606w}$) colour and the $V$-band mass-to-light ratio ($\log_{10}\Gamma_{V}$) for synthetic galaxies, with $\log_{10}(M/M_{\odot})>9.0$, extracted from {\sc galform} at $z=0$. The vertical lines denote the $15.87^{\rm th}$-$84.13^{\rm th}$ percentile range of distribution in several colour bins. The solid (green) line is the best-fitting linear relation between the colour and mass-to-light ratio. The coloured curves show the predicted tracks for exponentially decreasing star formation histories as in Fig. \ref{fig:theory}. The greyscale points show the location of a representative sub-sample of galaxies.}
\label{fig:s}
\end{figure}

As an example we consider the usefulness of the {\em intrinsic} observed-frame (B$_{f435w}-$V$_{f606w}$) colour as a diagnostic of the rest-frame $V$-band mass-to-light ratio at $z=0$. The (B$_{f435w}-$V$_{f606w}$) colour is chosen to demonstrate the usefulness of the technique as it represents the optimum choice of colour at $z=0$ (see Section \ref{sec:optimise}) with which to measure the mass-to-light ratio using the HST filter set. Comparing the {\em intrinsic} observed-frame (B$_{f435w}-$V$_{f606w}$) colour and the rest-frame $V$-band mass-to-light ratio ($\log_{10}\Gamma_{V}$) predicted by {\sc galform} at $z=0$, as shown in Fig. \ref{fig:s}, reveals the two quantities are highly correlated ($r\simeq 0.9$, where $r$ is Pearson's sample correlation coefficient\footnote{A value $|r_{xy}|=1.0$ implies the relationship between the variables $x$ and $y$ can be perfectly described by a linear equation, while $r=0.0$ on the other hand implies there is no linear correlation between the variables.}). Using linear regression, taking into account the abundance of each galaxy, we can fit the relationship between an arbitrary colour $(m_{a}-m_{b})$ and the $V$-band mass-to-light ratio by a linear relation, i.e.
\begin{eqnarray} \label{eqn:main}
\log_{10}\Gamma_{V} & = &\log_{10}[(M/M_{\odot})/(L_{V}/L_{V\odot})]\\
 & = &  p_{1}\times (m_{a}-m_{b}) + p_{2}.
\end{eqnarray} 
For the ($B_{f435w}-V_{f606w}$) colour / $V$-band mass-to-light ratio the parameters $p_{1}$ and $p_{2}$ are $1.10$ and $-0.50$ respectively and the linear fit is displayed in Fig. \ref{fig:s}. Reassuringly this linear fit is coincident with the predictions made in Section \ref{sec:theory} assuming simple exponentially decaying star formation histories (as shown in Fig. \ref{fig:theory} and Fig. \ref{fig:s}). 

\begin{figure}
\centering
\includegraphics[width=20pc]{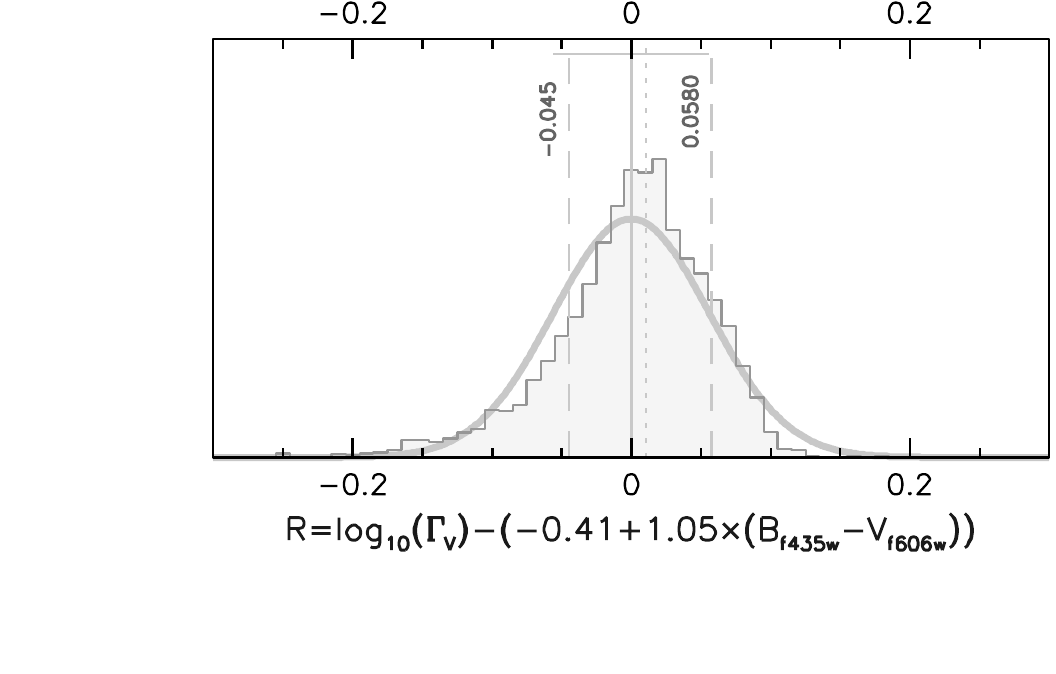}
\caption{The distribution of residuals ($R_{i}$), that is, the difference between the true mass-to-light ratio $\Gamma$ and that inferred from the intrinsic colour using the best-fit linear relation, Eq. \ref{eqn:main}. The solid grey curve represents a normal distribution with the same mean, standard deviation and area as $R_{i}$. The left and right vertical dashed lines denote the $15.87^{\rm th}$-$84.13^{\rm th}$ percentiles respectively, while the dotted line denotes the median.}
\label{fig:histo}
\end{figure}

\begin{figure}
\centering
\includegraphics[width=20pc]{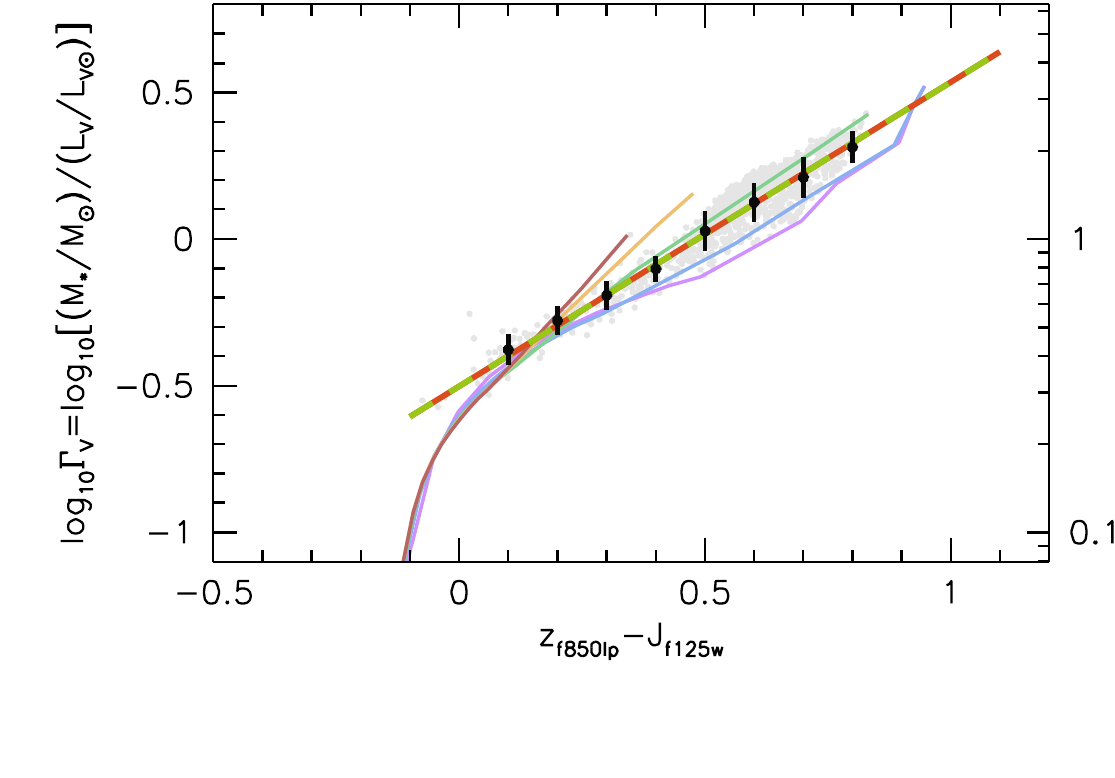}
\caption{Similar to Fig. \ref{fig:s} but showing instead the relationship between the ($z_{f850lp}-J_{f125w}$) and the $V$-band mass-to-light ratio ($\log_{10}\Gamma_{V}$) for synthetic galaxies, with $\log_{10}(M/M_{\odot})>9.0$, extracted from {\sc galform} at $z=1$.}
\label{fig:s2}
\end{figure}

If we now consider the {\em residual} distribution ($R$, shown in Fig. \ref{fig:histo}) between the mass-to-light ratio estimated from this linear relation and the true mass-to-light ratio, defined generally as,
\begin{equation}\label{eqn:fit}
R=\log_{10}\Gamma_{V} - (p_{1}(m_{a}-m_{b}) + p_{2}),
\end{equation} 
we find that, for the (B$_{f435w}-$V$_{f606w}$) colour, the standard deviation of $R$ is $\sigma_{\, R}\simeq 0.059$. This is a factor $\sim 4\times$ smaller than the standard deviation of the stellar mass-to-light ratio distribution ($\sigma_{lg\Gamma}=0.261$, see \S\ref{sec:galform.mtold}). The residual distribution (as shown in Fig. \ref{fig:histo}) has an approximately gaussian shape (Fig. \ref{fig:histo} also shows a normal distribution with the same mean and standard deviation as $R$) except for the tails of the distribution. This analysis suggests that a {\em perfectly} measured {\em intrinsic} colour can be used to estimate the mass-to-light ratio $\Gamma_{V}$ with a precision of $\simeq 14\%$. An additional example is shown in Fig. \ref{fig:s2}, where a similar relation is evident at $z=1$.  

\subsection{Sensitivity to stellar mass cut}\label{sec:mass_cut}

As noted in \S\ref{sec:sample_defn} our sample is effectively mass limited, with $\log_{10}(M_{*}/M_{\odot})>9$. By changing this criteria we can assess the sensitivity of the parameters $p_{1}$ and $p_{2}$. We find that both $p_{1}$ and $p_{2}$ vary by only a small amount ($p_{1}=1.10\to 1.07$, $p_{2}=-0.50\to -0.52$) as the mass limit is increased from $\log_{10}(M_{*}/M_{\odot})>9\to 11$.

\subsection{Dust}\label{sec:dust}

In the preceding analysis we demonstrated that the {\em intrinsic} colour of a galaxy provides a robust diagnostic of the stellar mass-to-light ratio. However, the {\em observed} SEDs of galaxies, and thus their colours and {\em observed} mass-to-light ratios are altered by dust, potentially affecting the reliability of a single colour as a diagnostic. Here we analyze the effects of dust assuming an attenuation law with a universal shape, with the overall amount of dust extinction specified by a single parameter which varies from galaxy to galaxy.

Over the UV, optical and NIR, dust acts to preferentially absorb light at shorter wavelengths. One effect of dust is then to {\em redden} the {\em observed} SED and thus the colour relative to the intrinsic, i.e. $(m_{a}-m_{b})^{\rm obs} = (m_{a}-m_{b})^{\rm int}+(A_{a}-A_{b})$, where $(A_{a}-A_{b})\ge 0.0$. However dust, by absorbing some fraction of the light, also acts to increase the {\em observed} mass-to-light ratio relative to the intrinsic value, (i.e. $\Gamma^{\rm\, obs}\ge\Gamma^{\rm\, int}$). Thus the effect of dust is to shift an individual galaxy in the colour-$\Gamma$ ratio plane by a vector ${\bf d}$ with orientation $0<\theta<\pi/2$ (with respect to the $\log_{10}\Gamma$ axis). That is, dust acts in {\em approximately} the same direction as the $(m_{a}-m_{b})$-$\log_{10}\Gamma_{V}$ relation.

The direction of this vector can be estimated analytically as follows: the relationship between the observed and intrinsic mass-to-light
ratio is,
\begin{eqnarray} \label{eqn:dust}
\log_{10}(\Gamma^{\rm\, obs}_{V}) &= &\log_{10}(\Gamma^{\rm\,
  int}_{V})+0.4\,E_{B-V}\,k_{V},
\end{eqnarray} 
where $E_{B-V}$ is the colour excess and $k_{i}$ is the effective reddening for a filter $i$ such that $k_{i}=A_{i}/E_{B-V}$. The extinction $A_{i}$ is defined as 
\begin{equation}
A_{i}=m^{\rm\, obs}_{i}-m^{\rm\, int}_{i}
\end{equation} 
which is in turn given by,
\begin{equation}
m^{\rm\, obs}_{i}-m^{\rm\, int}_{i} = -2.5\,\log_{10}Q,
\end{equation}
where,
\begin{equation}\label{eqn:dust2}
Q=\left(\int T\,f_{\nu}^{\rm int}\,10^{-0.4E_{B-V}k_{\lambda}}\,\frac{{\rm d}\nu}{\nu}\right)\left(\int T\,f_{\nu}^{\rm int}\,\frac{{\rm d}\nu}{\nu}\right),
\end{equation} 
where $T$ is the filter response curve and $f_{\nu}^{\rm int}$ is the intrinsic spectrum of the source\footnote{Examination of Eq. \ref{eqn:dust2} shows that $k_{i}$, for a given filter, is sensitive to the choice of $E_{B-V}$ and $f_{\nu}^{\rm int}$ and therefore differs for every distinct $f_{\nu}^{\rm int}$. However, over the range $E_{B-V}=0.0\to 2.0$ and for $f_{\nu}\propto\nu^{\beta+2}$, with $\beta=-2\to 2$, $k_{i}$ varies only at the percent level. A notable exception where this assumption  breaks down is when the light in a given filter is dominated by emission from a small wavelength interval; for example when the equivalent width of an emission line is comparable to the width of the filter, or where the filter itself stradles a spectral break.}. 

The relationship between between the observed and intrinsic colour is,
\begin{eqnarray} \label{eqn:dust}
(m_{a}-m_{b})^{\rm obs} & = &(m_{a}-m_{b})^{\rm int}+(A_{a}-A_{b})\\
& = & (m_{a}-m_{b})^{\rm int}+E_{B-V}(k_{a}-k_{b}).
\end{eqnarray} 
Thus, the dust vector is then given by,
\begin{equation}
{\bf d}=\left(E_{B-V}(k_{a}-k_{b}),0.4\,E_{B-V}\,k_{V}\right)
\end{equation}
  which has a gradient, 
\begin{equation}\label{eqn:d1}
d_{1}=0.4\times k_{V}/(k_{a}-k_{b}).
\end{equation} 

Using this formalism, and assuming a Calzetti et al. (2000) attenuation curve\footnote{The shapes of the commonly used dust attenuation curves, including the Calzetti et al. (2000) starburst, Milky Way, SMC and LMC (e.g. Pei et al. 1992) curves, are all similar in the rest-frame optical where the optimal colour is typically found.}, we determine that the gradient of the dust vector ${\bf d}$, for the (B$_{f435w}-$V$_{f606w}$) colour at $z=0$, is $d_{1}=1.30$ (c.f. $p_{1}=1.10$), i.e. while the $(m_{a}-m_{b})$-$\log_{10}\Gamma_{V}$ relation and dust vector point in approximately the same direction they are not perfectly aligned. Qualitatively the presence of a dust vector not perfectly aligned with the relation will then have two effects: it will both increase the scatter in the residual distribution (i.e. decrease the precision of the $(m_{a}-m_{b})$-$\log_{10}\Gamma_{V}$ relation) and introduce a systematic shift. 

The direction of the systematic shift depends on the angle between the dust vector gradient $d_{1}$ and $p_{1}$. For example, if the gradient of the dust vector is smaller than that of the intrinsic relation (i.e. $d_{1}<p_{1}$) the mass-to-light ratios inferred using the intinsic relation will be systematically too high, i.e. the stellar masses will be overestimated. The size of the systematic shift ($\delta\log_{10}\Gamma_{V}$) can be derived analytically,
\begin{equation}\label{eqn:d2} 
\delta\log_{10}\Gamma_{V}=(k_{a}-k_{b})(d_{1}-p_{1})E_{B-V}\equiv d_{2}\times E_{B-V}.
\end{equation}
For the (B$_{f435w}-$V$_{f606w}$) relation this gives $d_{2}=0.26$, suggesting only a small ($<0.1$) systematic shift for typical values of $E_{B-V}$ ($<0.3$). However, for extremely dusty galaxies the mass-to-light ratio will be overestimated. It is worth noting that the misalignment between ${\bf d}$ and the single-colour-$\Gamma$ relation in this case is in fact greater than every other optimum choice of colour (see the proceeding section). 

\subsubsection{The effect of normally distributed dust}

\begin{figure}
\centering
\includegraphics[width=20pc]{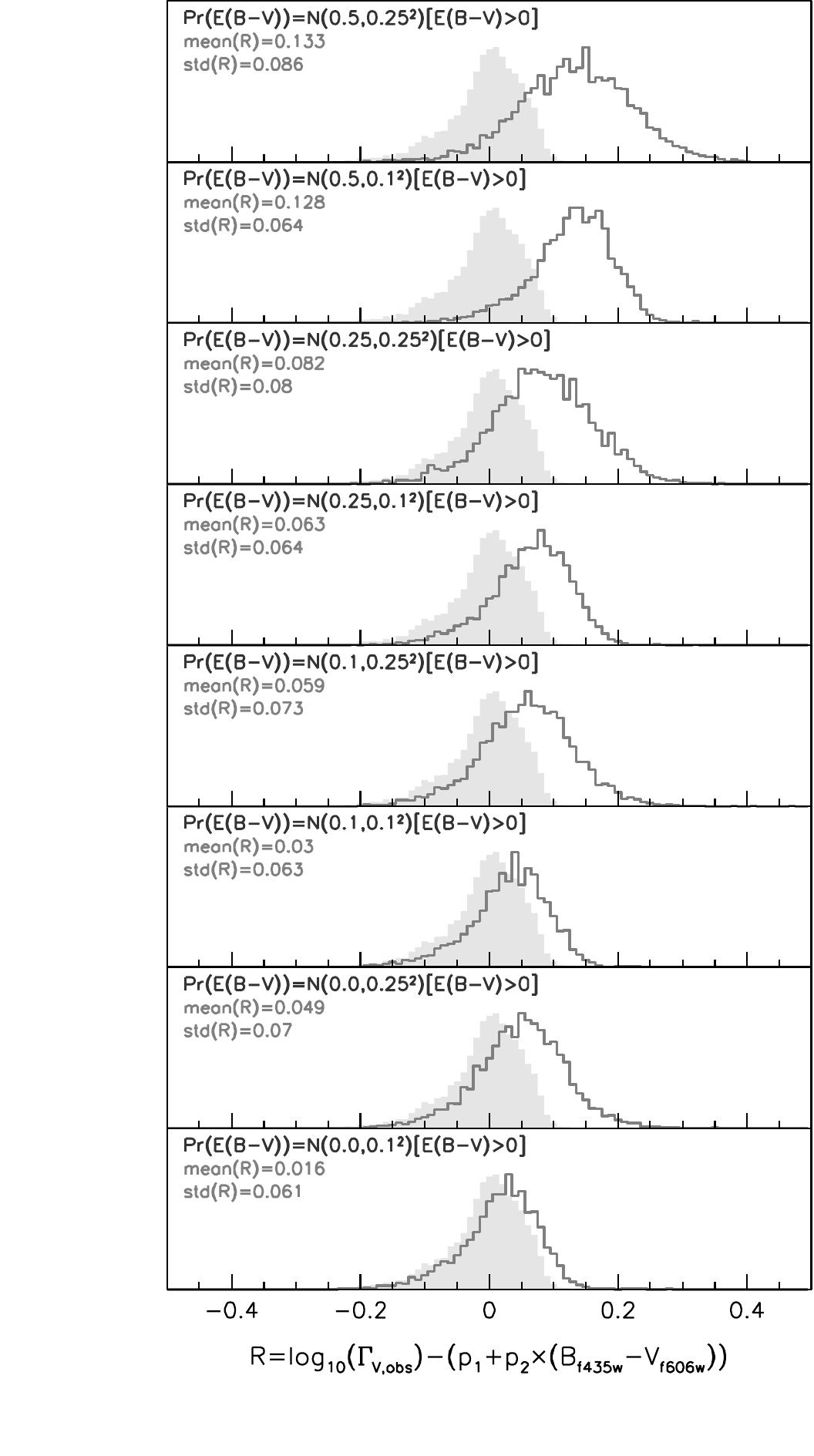}
\caption{The resulting distribution of stellar mass-to-light ratio residuals $R$ (Eq. \ref{eqn:fit}) after applying dust using $E_{B-V}$ drawn from a truncated normal distribution (Pr$(E_{B-V})=\mathcal{N}(\mu,\sigma^{2})[E_{B-V}\ge 0]$) for several different values of $\mu$ ($\in\{0.0,0.1,0.25,0.5\}$) and $\sigma$ ($\in\{0.1,0.25\}$). In all panels the shaded histogram shows the intrinsic distribution of residuals (as also shown in Fig. \ref{fig:histo}).}
\label{fig:dust}
\end{figure}

The effect of dust can be further illustrated by applying dust extinction to each galaxy according to some distribution of $E_{B-V}$. For this we assume each value of $E_{B-V}$ is drawn from a truncated normal distribution such that $E_{B-V}\ge 0.0$, i.e. Pr$(E_{B-V})=\mathcal{N}(\mu,\sigma^{2})[E_{B-V}\ge 0]$\footnote{Here $\mathcal{N}(\mu,\sigma^{2})$ denotes a normal distribution with mean $\mu$ and variance $\sigma^{2}$ while the term $[E_{B-V}\ge 0]$ forces the probability on un-physical negative values of $E_{B-V}$ to be zero using the Iverson bracket notation.}. Fig. \ref{fig:dust} demonstrates the effect of assuming this form of dust distribution on the {\em residual} distribution (for the B$_{f435w}-$V$_{f606w}$ relation at $z=0$) assuming various values of $\mu$ ($\in\{0.0,0.1,0.25,0.5\}$)\footnote{Because we have truncated the distribution $\mu$ is no longer equal to the mean/median value of $E_{B-V}$.} and $\sigma$ ($\in\{0.1,0.25\}$). Increasing $\mu=0\to 0.5$ causes a systematic increase in $R$ of $0.13$ (the same as the value predicted using Eq. \ref{eqn:d2}) while increasing the scatter $\sigma=0\to 0.25$ results in the standard deviation of $R$ increasing from $\approx 0.06$ to $\approx 0.09$.

\subsection{Photometric uncertainties}

Another advantage of a linear relation between the colour $(m_{a}-m_{b})$ and the stellar mass-to-light ratio $\log_{10}\Gamma_{V}$ is that it is easy to show that $\log_{10}\Gamma_{V}$ is gaussian assuming both the distribution of $R$ and the noise on the colour are also gaussian. The mean of $\log_{10}\Gamma_{V}$ is clearly $p_{1}(m_{a}-m_{b})+p_{2}$ and the uncertainty is then\footnote{This can be shown by integrating over the colour distribution.}:
\begin{equation}\label{eqn:uncertainty}
\sigma_{lg\Gamma}^{2}=p_{1}^{2}\sigma_{ab}^{2}+\sigma_{R}^{2},
\end{equation}
where $\sigma_{ab}$ is the uncertainty on the colour $(m_{a}-m_{b})$.

\section{Choosing the optimal colour}\label{sec:optimise}

We have now shown that even in the presence of dust a single observed colour provides a robust diagnostic of the stellar mass-to-light ratio. It is useful then to determine, for a series of redshifts ($0\le z\le 1.5$ for the HST filter set and $0\le z\le 0.5$ for the GAMA filter set), the optimum colour with which to estimate the mass-to-light ratio together with the associated parameters. When determining the optimum colour we consider all the possible combinations of filters in each of the filter sets described in \S\ref{sec:filters} (i.e. we consider 45 possible colours based on the $9$-filter HST set and 55 colours from the $10$-filter GAMA set). Colours probing the rest-frame UV typically provide the worst correspondence with the mass-to-light ratio ($R>0.13$, and in the case of the $FUV-NUV$ colour at $z=0$ $R\sim 0.3$). This is predominantly because the UV is strongly affected by the recent star formation history of the galaxy which does not necessarily correlate with overall mass-to-light ratio. At $z>0.3$ the FUV (and at $z>1.1$ the UV$_{f225w}$) band is also affected by the Lyman-limit break.

In all redshift cases a range of filter combinations produce roughly similar residual distributions, with $\sigma_{R}<0.07$. For example, at $z=0$ both the $($B$_{f435w}-$V$_{f606w})$ and $($B$_{f435w}-i_{f775w})$ yield $\sigma_{R}<0.06$. Because of this we can also dictate a preference for colours for which the difference between the $(m_{a}-m_{b})$-$\log_{10}\Gamma_{V}$ relation slope and the gradient of the dust vector ($d_{1}$) is minimised, thereby reducing the systematic shifts expected due to dust. 

For both filter sets and a range of redshifts (GAMA: $z=0.0-0.5$, HST: $z=0.0-1.5$) Table \ref{tab:params} gives the best-fitting colour and gives the parameters $p_{1}$, $p_{2}$, $d_{1}$ and $d_{2}$ for each redshift. In virtually all cases the optimal choice of colour covers the rest-frame $0.4-0.7\mu {\rm m}$ range. In each case the standard deviation of the relevant {\em intrinsic} residual distribution is $\sigma_{R}<0.07$. In most cases the dust vector is closely aligned (typically $|d_{2}|<0.15$) with the $(m_{a}-m_{b})$-$\log_{10}\Gamma_{V}$ relation with the greatest (fractional) offset being the (B$_{f435w}-$V$_{f606w}$) colour at $z=0$ (considered in the previous section).

Thus far we have derived the stellar mass-to-light ratio - single colour relation using the rest-frame $V$-band luminosity. To determine the stellar mass it is then necessary to apply a $k$-correction to an observed luminosity (to determine the rest-frame $V$-band luminosity) introducing an additional source of uncertainty. However, we can instead define the mass-to-light ratio - single colour relation using an observed frame luminosity thereby removing the need for a $k$-correction. This is detailed in Appendix \ref{app:kcor}.

\begin{table*}
\caption{The optimum {\em observed}-frame colour for estimating the stellar mass-to-light ratio for various redshifts assuming the GAMA ({\em top}) and HST ({\em bottom}) filter sets. $^{1}$ the observed frame colour chosen to measure the mass-to-light ratio. $^{2}$ the rest-frame wavelength range. $^{2}$ Pearson's correlation coefficient between $\log_{10}\Gamma_{V}$ and the colour $(m_{a}-m_{b})$. $^{3}$ The standard deviation of the {\em intrinsic} residual distribution. $^{4}$ Parameters describing the best fit using simple linear regression (see Equation \ref{eqn:fit}). $^{5}$ The gradient of the dust vector ${\bf d}$ (defined in Eq. \ref{eqn:d1}). $^{6}$ The factor relating the colour excess $E_{B-V}$ and the systematic offset to $\log_{10}\Gamma$ (defined in Eq. \ref{eqn:d2}).}
\begin{tabular}{ccccccccccc}
\hline
 $z$ & $(a-b)$$^{1}$ & $|r|$$^{2}$ & $\sigma_{R}$$^{3}$ & $p_{1}$$^{4}$ & $p_{2}$$^{4}$ & $d_{1}$$^{5}$ & $d_{2}$$^{6}$\\
\hline
\multicolumn{8}{c}{GAMA} \\
\hline
0.0 & $g-r$ & 0.91 & 0.056 & 1.40 & -0.49 & 1.29 & -0.13   \\
0.1 & $g-i$ & 0.92 & 0.056 & 0.80 & -0.45 & 0.78 & -0.04   \\
0.2 & $g-z$ & 0.90 & 0.060 & 0.59 & -0.57 & 0.58 & -0.03   \\
0.3 & $r-z$ & 0.94 & 0.053 & 1.09 & -0.41 & 1.07 & -0.02   \\
0.4 & $r-Y$ & 0.94 & 0.054 & 0.75 & -0.44 & 0.75 & -0.01   \\
\hline
\multicolumn{8}{c}{HST} \\
\hline
0.0 & B$_{f435w}-$V$_{f606w}$ & 0.90 & 0.059 & 1.10 & -0.50 & 1.30 & ~0.25   \\
0.25 & V$_{f606w}-i_{f775w}$ & 0.93 & 0.058 & 1.26 & -0.41 & 1.24 & -0.04   \\
0.5 & $i_{f775w}-z_{f850lp}$ & 0.95 & 0.05 & 2.37 & -0.51 & 2.4 & ~0.02   \\
0.75 & $i_{f775w}-$J$_{f125w}$ & 0.95 & 0.051 & 0.8 & -0.54 & 0.81 & ~0.02   \\
1.0 & $z_{f850lp}-$J$_{f125w}$ & 0.95 & 0.054 & 1.04 & -0.5 & 1.19 & ~0.21   \\
1.25 & Y$_{f105w}-$H$_{f160w}$ & 0.96 & 0.05 & 0.94 & -0.53 & 0.98 & ~0.07   \\
1.5 & J$_{f125w}-$H$_{f160w}$ & 0.97 & 0.051 & 1.7 & -0.57 & 1.75 & ~0.05   \\
\hline
\end{tabular}
\label{tab:params}
\end{table*}

\section{Comparison with other studies}\label{sec:sed_fitting}

\subsection{Other determinations of single-colour relations}

It is interesting to assess whether our relation(s) are similar to those derived empirically in the literature. Taylor et al. (2011, hereafter T11) define an empirical relation using stellar masses and ($k$-corrected) colours from the Galaxy and Mass Assembly (GAMA) spectroscopic survey (Driver et al. 2011). The GAMA survey {\em currently} incorporates UV, optical and near-IR photometry from GALEX, SDSS and UKIDSS respectively with spectroscopy coming from a variety of sources (predominantly from SDSS and AAO). T11 derives stellar masses using a Bayesian SED fitting methodology using only optical (SDSS $ugriz$) photometry. The T11 single-colour - mass-to-light ratio relation is defined using the rest-frame $(g-i)$ colour and $i$-band luminosity such that,
\begin{equation}\label{eqn:T12}
\log_{10}\Gamma_{i,ab}=0.7\times (g-i)_{r}+1.15,
\end{equation}
where $\log_{10}\Gamma_{i,ab}$ is defined as the mass-to-light ratio using AB centric units (i.e. luminosity has units of $L_{i}(M_{i}=0.0)$).

At $z=0$ we find that the $(g-r)$ colour is favoured over $(g-i)$ due to the slightly smaller value of $R$. However, we can nevertheless easily define a relation between the $(g-i)$ colour and and $i$-band luminosity (at $z=0$), finding:
\begin{equation}\label{eqn:T12_ours}
\log_{10}\Gamma_{i,ab}=0.80\times (g-i)_{r}+1.41,
\end{equation}
where $\log_{10}\Gamma_{i,ab}$ is defined using the same AB centric unit system in Eq. \ref{eqn:T12}.

Clearly there is some disagreement between this relation and the T11 relation. However there are several important differences in their construction which need to be addressed. Firstly, while our relation is defined using a sample of galaxies at $z=0$ (such that the observed $g-i$ colour is also the rest-frame $g-i$ colour) the T11 (GAMA) sample encompasses a range of redshifts ($z<0.65$; median $z=0.2$). If instead we consider the rest-frame $(g-i)$ colour at the median GAMA redshift ($z=0.2$) we find,
\begin{equation}\label{eqn:T12_ours_rest}
\log_{10}\Gamma_{i,ab}=0.74\times (g-i)_{r}+1.41,
\end{equation}
the slope of which is now consistent with that derived by T11. Secondly, stellar masses in T11 were derived assuming a Chabrier (2003) initial mass function (IMF) while, in our case, {\sc galform} was run assuming a Salpeter (1955) IMF. As noted in the introduction stellar masses estimated assuming a Salpeter IMF are typically around $0.2\,{\rm dex}$ larger than those assuming the Chabrier (2003) IMF. Making this correction to the normalisation of the T11 relation (i.e. $1.15+0.2=1.35$) then yields good agreement between this single colour mass-to-light ratio relation and that of T11.

\subsection{Multi-wavelength broadband SED fitting}

It is also useful to comment on the precision of a single-colour diagnostic compared with typical broad-band SED fitting techniques. Several recent studies have attempted to determine the accuracy and precision ({\em uncertainty}) with which broad-band SED modelling can recover various physical properties, including the stellar mass-to-light ratio (e.g. Pacifici et al. 2012, Pforr et al. 2012, Wilkins et al. \tr{{\em in prep}}). 

Wilkins et al. \tr{{\em in prep}} uses the same library of synthetic galaxies considered in this paper to investigate the the accuracy and precision (uncertainty) with which the stellar mass-to-light ratio can be recovered for various observational factors (including the availability of {\em a priori} redshifts, rest-frame SED coverage and S/N), modelling choices (including star formation history parameterisation, statistical indicator) and as a function of various galaxy properties (including stellar mass, mass-to-light ratio, star formation rate, redshift). Wilkins et al. \tr{{\em in prep}} uses a Markov Chain Monte-Carlo likelihood exploration algorithm with an adaptive Metropolis proposal using the {\sc PyMC} package\footnote{https://github.com/pymc-devs/pymc} ({\sc BayesME}, Zuntz \& Wilkins {\em in prep}) assuming, for the most part, a library built using simple exponentially decaying star formation histories ($\psi\propto e^{-t/\tau}$) and a fixed metallicity. In the {\em default} scenario of Wilkins et al. \tr{{\em in prep}} (synthetic galaxies at $z=0.5$ with the redshift known {\em a priori}, broadband observed frame UV-optical-NIR photometry with uniform noise where S/N($H_{f160w}$)$=100$, the $V$-band mass-to-light ratio was recovered with an accuracy of $0.05$ (where the accuracy is defined as the median of the {\em residual} distribution $R$, where $R=\langle \log_{10}\Gamma_{V}\rangle-\log_{10}\Gamma_{V}^{\rm true}$), and uncertainty $0.05$ (where the this is defined as half of the $15.9^{\rm th}$-$84.1^{\rm th}$ percentile range of the {\em residual} distribution). Decreasing the number of observed bands, decreasing the S/N or increasing the redshift all result in the uncertainty growing. 

The optimal colour relation at $z=0.5$ for the HST filter set is $i_{f775w}-z_{f850lp}$ with $p_{1}=2.37$, $p_{2}=-0.51$ and $\sigma_{R}=0.05$. A S/N of $100$ in both $i_{f775w}$ and $z_{f850lp}$ suggests $\sigma_{gr}^{2}=2\times (1.0875/100)^{2}\simeq 0.00012$ and thus the uncertainty on $\sigma_{lg\Gamma}$, in the absence of dust is, using Eq. \ref{eqn:uncertainty}, $\sigma_{lg\Gamma}\simeq 0.06$. Degrading the S/N to $30$ yields $\sigma_{lg\Gamma}\simeq 0.10$. 

It thus appears that a single-colour diagnostic is capable, despite the availability of significantly less information, of recovering the stellar mass-to-light ratio with a similar uncertainty to this implementation of SED fitting. While this may seem counter-intuitive it reflects the fact that in this case improving the sophistication of the fitted model does more for the quality of the results than simply adding more data.

A simple intuitive example of this would be an implementation of SED fitting which assumes all galaxies have a constant star formation history. Such a model is clearly not a good description of the real Universe given observations of passively evolving galaxies. As such the model is unlikely to provide an accurate estimate of the mass-to-light ratio (of passively evolving galaxies). In such a case adding more data is unlikely to improve the accuracy/precision of the fit and a single-colour diagnostic (based on a more sophisticated model) would unambiguously provide a more accurate diagnostic of the stellar mass-to-light ratio.  

\section{Conclusions}\label{sec:c}

Using the {\sc galform} galaxy formation model to produce realistic galaxy star formation and metal enrichment histories we have investigated the correlation between the stellar mass-to-light ratio and various {\em observed} frame colours.

We find that several {\em observed}-frame colours (typically within the {\em rest}-frame optical) and the $V$-band stellar mass-to-light ratio are often highly correlated ($|r|>0.8$, where $r$ is Pearson's correlation coefficient) and that the ($V$-band) mass-to-light ratio can be predicted from the {\em intrinsic} colours, using a simple linear relation ($\log_{10}\Gamma_{V}=p_{1}(m_{a}-m_{b}) + p_{2}$), with an uncertainty of $\sigma_{lg\Gamma} < 0.07$. While the addition of dust introduces both a systematic offset and increases the uncertainty, if the choice of filters is made carefully these effects are typically small. 

The uncertainties predicted to arise using a single-colour diagnostic ($\sigma_{lg\Gamma}=0.05-0.15$) are comparable to those obtained using a wider set of observations but simplistic star formation and metal enrichment histories (e.g. exponentially decaying star formation histories).

For future use we provide details of the optimal choice of colour (for two common sets of filters), along with the associated parameters necessary to accurately determine the stellar mass-to-light ratio at $z=0\to 1.5$.

\subsection*{Acknowledgements}

{\sc galform} was run on the ICC Cosmology Machine, which is part of the DiRAC Facility jointly funded by STFC, the Large Facilities Capital Fund of BIS, and Durham University. SMW acknowledges support from STFC. VGP acknowledges support from the UK Space Agency. VGP, CGL and CMB acknowledge support from the Durham STFC rolling grant in theoretical cosmology. JZ acknowledges support from ERC starting grant 240672 and a James Martin fellowship.

\bsp

\appendix
\section{observed frame mass-to-light ratio}\label{app:kcor}

To avoid the need for a $k$-correction (from an observed frame luminosity to the rest-frame $V$-band) when estimating stellar masses (as opposed to the stellar mass-to-light ratio) we also provide parameters relating the observed frame colour to the observed frame mass-to-light ratio, i.e.: 
\begin{eqnarray} \label{eqn:main_obs}
\log_{10}\Gamma_{b,ab} & = &\log_{10}[(M/M_{\odot})/(L_{b}/L_{b}(M_{b}=0.0))]\\
 & = &  p_{1}\times (m_{a}-m_{b}) + p_{2}.
\end{eqnarray} 
For each redshift/colour the redder filter is chosen as the reference band for the luminosity ($L_{b}$) as there is typically less variation in the mass-to-light ratio. The units of luminosity are chosen to be AB centric (as opposed to the more familiar solar units in the case of rest-frame $V$-band mass-to-light ratio) and are defined as the luminosity at $M_{b}=0.0$ (i.e. $L_{b}(M_{b}=0.0)$). With the exception of the parameter $d_{1}$ (defined in Eq. \ref{eqn:d1}), which has to be re-defined as
\begin{equation}\label{eqn:d1_obs}
d_{1}=0.4\times k_{b}/(k_{a}-k_{b}),
\end{equation} 
(i.e. replacing the $k_{V}$ term with $k_{b}$) we can derive the same set of parameters ($r$, $\sigma_{R}$, $p_{1}$, $p_{2}$, $d_{2}$) that were used to describe the rest-frame $V$-band mass-to-light ratio - single colour relation for the observed frame relation. These parameters are given in Table \ref{tab:params_obs}.

\begin{table*}
\caption{Parameters relating the {\em observed}-frame colour $(m_{a}-m_{b})$ with the observed frame mass-to-light ratio for the same colours, redshifts, and filter sets in Table \ref{tab:params}. The observed frame reference band is chosen to be redder filter of each colour (i.e. the $b$-band). $^{1}$ the observed frame colour chosen to measure the mass-to-light ratio. $^{2}$ Pearson's correlation coefficient between $\log_{10}\Gamma_{b,ab}$ and the colour $(m_{a}-m_{b})$. $^{3}$ The standard deviation of the {\em intrinsic} residual distribution. $^{4}$ Parameters describing the best fit using simple linear regression (see Equation \ref{eqn:main_obs}). $^{5}$ The gradient of the dust vector ${\bf d}$ (re-defined in Eq. \ref{eqn:d1_obs} for the observed frame $b$-band ). $^{6}$ The factor relating the colour excess $E_{B-V}$ and the systematic offset to $\log_{10}\Gamma$ (defined in Eq. \ref{eqn:d2}).}
\begin{tabular}{ccccccccccc}
\hline
 $z$ & $(a-b)$$^{1}$ & $|r|$$^{2}$ & $\sigma_{R}$$^{3}$ & $p_{1}$$^{4}$ & $p_{2}$$^{4}$ & $d_{1}$$^{5}$ & $d_{2}$$^{6}$\\
\hline
\multicolumn{8}{c}{GAMA} \\
\hline
0.0 & $g-r$ & 0.89 & 0.056 & 1.24 & 1.45 & 1.12 & -0.15   \\
0.1 & $g-i$ & 0.88 & 0.057 & 0.69 & 1.35 & 0.60 & -0.19 \\
0.2 & $g-z$ & 0.86 & 0.061 & 0.63 & 1.20 & 0.64 & ~0.02   \\
0.3 & $r-z$ & 0.91 & 0.056 & 0.92 & 1.31 & 0.82 & -0.16   \\
0.4 & $r-Y$ & 0.91 & 0.052 & 2.31 & 1.22 & 1.96 & -0.25   \\
\hline
\multicolumn{8}{c}{HST} \\
\hline
0.0 & B$_{f435w}-$V$_{f606w}$ & $0.37-0.71$$\mu {\rm m}$ & 0.89 & 0.059 & 1.02 & 1.43 & 1.24 & ~0.28 \\
0.25 & V$_{f606w}-i_{f775w}$ & $0.37-0.69$$\mu {\rm m}$ & 0.91 & 0.062 & 1.16 & 1.34 & 1.08 & -0.11 \\
0.5 & $i_{f775w}-z_{f850lp}$ & $0.46-0.69$$\mu {\rm m}$ & 0.93 & 0.052 & 2.20 & 1.17 & 2.14 & -0.04 \\
0.75 & $i_{f775w}-$J$_{f125w}$ & $0.39-0.80$$\mu {\rm m}$ & 0.91 & 0.054 & 0.68 & 1.06 & 0.60 & -0.16 \\
1.0 & $z_{f850lp}-$J$_{f125w}$ & $0.41-0.70$$\mu m$ & 0.93 & 0.065 & 1.91 & 1.09 & 1.92 & ~0.00   \\
1.25 & Y$_{f105w}-$H$_{f160w}$ & $0.40-0.75$$\mu m$ & 0.94 & 0.052 & 0.80 & 0.96 & 0.76 & -0.07   \\
1.5 & J$_{f125w}-$H$_{f160w}$ & $0.44-0.68$$\mu m$ & 0.96 & 0.046 & 1.59 & 0.88 & 1.52 & -0.06   \\
\hline
\end{tabular}
\label{tab:params_obs}
\end{table*}

\end{document}